\documentclass{article}

\usepackage{arxiv}

\usepackage[utf8]{inputenc} 
\usepackage[T1]{fontenc}    
\usepackage{hyperref}       
\usepackage{url}            
\usepackage{booktabs}       
\usepackage{amsfonts}       
\usepackage{nicefrac}       
\usepackage{microtype}      
\usepackage{graphicx}
\usepackage[square,numbers]{natbib}
\usepackage{doi}
\usepackage{tabularx}
\usepackage{rotating}
\usepackage{pdflscape}
\usepackage{longtable}

\title{Satellite Image and Machine Learning based Knowledge Extraction in the Poverty and Welfare Domain}

\author{ \href{https://orcid.org/0000-0000-0000-0000}{\includegraphics[scale=0.06]{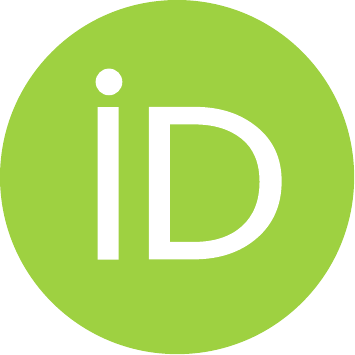}\hspace{1mm}Ola Hall}\thanks{Corresponding author.} \\
	Department of Human Geography\\
	Lund University, Sweden\\
	\texttt{ola.hall@keg.lu.se} \\
    \And
	\href{https://orcid.org/0000-0000-0000-0000}{\includegraphics[scale=0.06]{orcid.pdf}\hspace{1mm}Mattias Ohlsson} \\
	Center for Applied Intelligent Systems Research\\
	Halmstad University, Sweden\\
	\texttt{mattias.ohlsson@hh.se}
	\And
	\href{https://orcid.org/0000-0000-0000-0000}{\includegraphics[scale=0.06]{orcid.pdf}\hspace{1mm}Thorsteinn Rögnvaldsson} \\
	Center for Applied Intelligent Systems Research\\
	Halmstad University, Sweden\\
	\texttt{thorsteinn.rognvaldsson@hh.se} \\
}

\renewcommand{\shorttitle}{\textit{arXiv} Template}

\hypersetup{
pdftitle={A template for the arxiv style},
pdfsubject={q-bio.NC, q-bio.QM},
pdfauthor={Ola Hall, Mattias Ohlsson, Thorsteinn Rögnvaldsson},
pdfkeywords={First keyword, Second keyword, More},
}

\begin{document}
\maketitle

\begin{abstract}
Recent advances in artificial intelligence and machine learning have created a step change in how to measure human development indicators, in particular asset based poverty. The combination of satellite imagery and machine learning has the capability to estimate poverty at a level similar to what is achieved with workhorse methods such as face-to-face interviews and household surveys. An increasingly important issue beyond static estimations is whether this technology can contribute to scientific discovery and consequently new knowledge in the poverty and welfare domain. A foundation for achieving scientific insights is domain knowledge, which in turn translates into explainability and scientific consistency. We review the literature focusing on three core elements relevant in this context: transparency, interpretability, and explainability and investigate how they relates to the poverty, machine learning and satellite imagery nexus. Our review of the field shows that the status of the three core elements of explainable machine learning (transparency, interpretability and domain knowledge) is varied and does not completely fulfill the requirements set up for scientific insights and discoveries. We argue that explainability is essential to support wider dissemination and acceptance of this research, and explainability means more than just interpretability.
\end{abstract}

\keywords{Poverty \and Satellite imagery \and Machine Learning}

\section{Introduction}
A longstanding and to date contested theoretical and methodological question in the social sciences and economics is how to measure and monitor changes in poverty and human development over time \citep{Gibson2016} \citep{Ravallion2020}. UN has formed poverty reduction goals and called for a data revolution to support the fulfillment of those. The importance of the need to adequately measure and monitor poverty cannot be over emphasised, thus ”the world’s population should be counted, measured, weighted, and evaluated.” \citep{Jerven2017} p. 31. The traditional tools for gathering data on human development, face-to-face interviews and household surveys, are now supplemented with digital data sources and tools. Central to the “data revolution” has been the growing abundance and quality of satellite data together with recent developments in machine learning (ML), in particular deep learning and convolutional neural networks. For some human outcome indicators such as asset-based poverty, ML-approaches are now matching the performance of survey data.

From a policy or decision-making perspective, there is a growing interest in these new methods but downstream applications are rare \citep{BurkeEtAl2021} \citep{Blumenstock2020}. The limited adoption is probably driven by several forces, but it is evidenced from the literature that so far work has focused on performance aspects of the technology, i.e. improving prediction metrics. This is natural, considering the recency of the technology. Consequently, stakeholder interests are largely set aside resulting in weak ties to the larger issues in human development research, for example, understanding and combating poverty. 

The geography of poverty is reasonably well known, keeping in mind that traditional surveys rest on sample frameworks making them representative mainly at national, sometimes regional, scales, leaving many areas uncharted (e.g., Demographic and Health Surveys). Here, satellite imagery and ML certainly play an important role augmenting existing efforts and increasing the resolution of the sample grid. \textit{Why} people are poor is a more complicated question to answer and relates to the multi-dimensional nature of poverty and the uneven geographical distribution of both wealth and extreme poverty but also spatial patterns where extreme poverty and affluence are interspersed in the same locations. One approach from the toolbox of social scientists to unravel the causes and studying welfare and poverty is with case studies. Östberg et al. \citep{Ostberg2018} provide a good example of a longitudinal localized study in Tanzania. They combined local wealth rankings (e.g., counting tin roofs, size of farms) with interviews on livelihood matters and concluded that improved infrastructure and local entrepreneurs played an important role in the area’s improvement over time.

A pertinent question is if these new ML tools can be used to provide scientific insight on a a large scale into the question of why people are poor. Can they provide more basis for knowledge than ``just'' an estimate of the poverty? An example to illustrate this is the Cobb Douglas production function commonly found in economics of small firms but also in agricultural production (from \citep{neumann2010yield}), 
\[
\ln{(q)} = \sum_i \beta_i \ln{(X_i)} + v - u \;,
\]
where $\ln{(q)}$ is the logarithm of the production of the grid cell in question, $X_i$ are the different production inputs for that cell, $\beta_i$ are unknown parameters to be estimated and $v$ is a random error to account for statistical noise. Further deviations are due to inefficiencies $(u)$ for that grid cell. The function for crop production $(cp)$ could be
\[
\ln{(cp)} = \beta_0 + \beta_1\ln{(\mbox{\it temperature})} + \beta_2\ln{(\mbox{\it precipitation})} + \beta_3\ln{(\mbox{\it par})} + \beta_4\ln{(\mbox{\it soilfertility})} +v - u  \;,
\]
with the most important growth defining factors according to theory inserted; {\it temperature, precipitation, photo-synthetically active radiation} (\textit{par}), and a {\it soil fertility constant} (all values estimated for the particular grid cell studied). For the inefficiency $(u)$, influences of {\it land management, labor force, general accessibility} and {\it market access} are considered important: 
\[
u = \delta_1 (\mbox{\it irrigation}) + \delta_2 (\mbox{\it slope}) + \delta_3 (\mbox{\it agripopulation}) + \delta_4 (\mbox{\it access}) + \delta_5 (\mbox{\it market}).
\]
A model of this kind provides grounds for evaluating the effect of changes in explanatory variables on $\ln{(q)}$ and deviations from expected levels (the inefficiency function). It would be relevant to extract a similar explanation from the ML and satellite image based models for poverty estimation. What features in the satellite image form the basis for the poverty estimate? Is there an expected base level and place specific inefficiencies relating to the particular grid cell?
%
It is our position that future advances in this research domain must go together with better interpretability (what features in an image causes a certain score) and explainability (how do changes in one feature affect other features and ultimately the target) to be considered useful outside narrow groups. The aim of this paper is to investigate state-of-the-art of explainable AI (XAI) in the poverty and welfare domain of machine learning and satellite imagery.

\subsection{Remote sensing and machine learning predictions of poverty}
In the mid-1990s the National Aeronautics and Space Administration (NASA) approached the research community in an effort to realize the potential of satellite imagery – specifically addressing the social sciences. High hopes were expressed in People and Pixels: Linking Remote Sensing and Social Science \citep{National1998}. But the results have been meager and their added-value questioned \citep{Hall2010} \citep{Longley2002} until recently. Without comparison the bulk of work where satellite data is applied to social science relevant problems originates from the Defense Meteorological Program (DMSP) Operational Line-Scan System (OLS) after scholars already in the mid-1970’s observed that imagery showed the extent and intensity of human settlements. Data with this capability is usually referred to as nighttime lights (NTL). Technical limitations with data storage and processing power hampered the accessibility of imagery and development until the 1990’s. 

Early work observed that regions emitting high levels of NTL also were associated with high levels of economic output \citep{Elvidge1997}. Henderson and others used data on annual global nighttime lights and showed the linkage between artificial light and economic activity \citep{Henderson2012} \citep{Chen2011} \citep{Keola2015} \citep{Noor2008}. While the relation was observed much earlier by \citep{Elvidge1997} the previous authors detailed the econometrics, a work that was continued with \citep{Mellander2015} that showed that such data correlates closely with detailed records of wage income in Sweden. NTL data has been applied and evaluated in some of the poorest regions of the world. The work of \citep{Noor2008} showed that NTL correlated with asset-based measures of wealth for 37 countries in Africa but at the same time underlined the observation that nighttime lights underperform in the poorest regions, which was confirmed in a study in Burkina Faso \citep{Hall2019}. 
 
These limitations inspired recent papers that use daytime satellite imagery to measure poverty in developing countries, particularly in SSA. Pioneering works by Xie \citep{XieEtAl2016} and Jean et al. \citep{JeanEtAl2016} combined the power of NTL with daytime satellite imagery and recent tools in machine learning. To circumvent the lack of labeled training data they applied a two-step transfer learning approach to five countries in sub-Saharan Africa using NTL as labels, improving R2 by more than 10\%. This approach has been elaborated further in several papers and for countries outside Africa, for example, Sri Lanka, China and India, and with steadily improving performance metrics \citep{BurkeEtAl2021}. The workhorse indicator for these studies is the wealth index (WI) from the Demographic and Health Survey (DHS). The index is a principal components analysis of items that are easily observable from the surveying officer’s perspective, access to water, phones and bicycles, for example. On the other hand, Head et al. \citep{HeadEtAl2017} have shown that this method does not generalize in the same way other measures of development predict access to drinking water and a variety of health indicators. Other measures such as the consumption index from the living standards measurement survey (LSMS) are not predictive at the same level as asset-based indices. Overall, satellite image-based poverty predictions can now explain more than half of the variation and sometimes up to 85\% of the survey-measured poverty. 

\subsection{Measuring poverty}
There are two main approaches to measuring poverty. The first is absolute measures that use poverty lines with constant real value as in the World Bank definition of extreme poverty (i.e. those who live on less than \$1.90 a day). The second uses relative measures for which the poverty line varies as a function of a set proportion in relation to current mean (or median) \citep{Ravallion2020}. Examples of both can be found in the cited literature.

There is a fundamental difference between stock measures (assets) and flow measures (income, consumption, expenditures). While it is desirable to have data on household income and expenditures, they are limited by factors such as seasonality, misreporting and volatility \citep{RutsteinS2014}. Household assets are easier to collect as they involve items that are easily observed by the surveying officer and are closely linked to long-term welfare status. The majority of studies use asset wealth as an indicator of poverty. Several attempt to estimate consumption expenditure but with less success. There are also attempts to estimate the global multidimensional poverty index and also poverty rates (Table 4).

The wealth index (WI) used in most of the literature is the DHS WI which is a composite measure of a household's cumulative living standard. The WI is calculated based on a household’s ownership of selected assets, such as televisions, bicycles, materials used for housing construction (flooring and tiling), and types of water access and sanitation facilities. The WI places individual households on a continuous scale of relative wealth. DHS separates all interviewed households into five wealth quintiles to compare the influence of wealth on various population, health and nutrition indicators. Later versions also include land holdings and farm animals. 

The WI focus on tangible and visible physical objects is a requirement from a satellite image perspective. Although, several collected assets are normally located inside houses thus excluded from satellite surveying  roofing material, vehicles, water pumps, electrification and several others are, provided sufficient spatial resolution, accessible from above. Indeed, the literature is not clear on the relation between image features and which surveyed features that impact output predictions with two exceptions \citep{AyushEtAl2020} and \citep{AyushEtAl2021}. It is plausible that the ML-model is also picking up on some other features (besides the observable DHS assets) present in the image such as size and shape of arable land or road quality, properties that are known to be associated with welfare status. Complicating the matter further, we also know that there is evidence that much persistent poverty is place-based and geographically determined \citep{YehEtAl2020a} meaning that there might be associations between landscape specific properties and poverty that the model is susceptible to. For our concerns, sorting and ranking among impact features are probably the single most important future research direction. Herein lies the possibility for understanding something new about poverty and its determinants. 

The "poor" include both those who are always poor and those who move in and out of poverty. Baulch and Hoddinott \citep{Baulch2000} found that 20-65\% from 13 panel studies were classified as sometimes poor which were more numerous than the always poor category. McBride \citep{McbrideEtAl2021} means that being able to distinguish between structurally poor and stochastically poor is crucial for well-targeted interventions and would require models (for targeting, mapping, monitoring or early warning) to account for the structural determinants of impoverishment. This relates to the potential to observe if change occurs in either positive or negative direction. It is noted that very few studies address questions about change and how to measure change \citep{BurkeEtAl2021}. While DHS assets are meant to represent long-term welfare and surveys to be repeated at least five years apart, it is not evident how this methodology can capture change in poverty. Add that some forms of poverty are place-based and geographically determined and it is likely that the frameworks at hand are static in this respect. 

Places and regions with high levels of permanent poverty are more impacted by crises so understanding the patterns of poverty is a key task. Geographic targeting based on poverty maps are highly useful for identifying the vulnerable areas with geographic aggregates showing the spatial distribution of welfare. Targeting is then directed at identifying the households and individuals at risk. McBride \citep{McbrideEtAl2021} discusses the trade-offs between predictive models with hundreds of features and one that includes a few well-known easily observed variables. The latter being useful for poverty targeting, evaluation and monitoring and the former more suited for poverty mapping. To avoid replicating well-known poverty patterns it is important that the two converge. 

\section{Explainable AI}
There have recently been several surveys on XAI methods and terminology. Most relevant for our discussion is the one by Roscher et al. \citep{RoscherEtAl2020} where they discuss requirements for using machine learning (ML) for scientific discovery and organize them into three core elements.

They make a useful distinction between \textit{transparency}, \textit{interpretability}, and \textit{explainability}, where transparency considers the ML approach, interpretability considers the ML model together with data, and explainability considers the model, the data, and human involvement. 

\paragraph{Transparency}
In general ML models are transparent in the to the extent that they are mathematically well defined; equations can be written down that describe what they do. However, ML models tend not to be transparent in the sense that it is easy to understand why certain model design choices were made (e.g. number of layers, activation functions, regularization, training algorithm, etc.). Transparency relates to the processes for constructing the ML model, e.g., the final model itself, methods for model structure choices, and for fitting the parameters. If these can all be well described and motivated then the ML model is transparent. Following Lipton \citep{Lipton2018}, Roscher et al. \citep{RoscherEtAl2020} divide transparency into three parts: model transparency, design transparency, and algorithmic transparency. 

Model transparency relates to the transperancy of the structure of the model (e.g. the number of layers, activation functions, kernel functions, number of decision trees in a random forest, splitting criteria, etc.). Design transparency refers to design choices made when constructing the ML algorithm, are those choices understandable, well-motivated, and replicable? Examples are selecting neural network architecture, activation functions, training time, batch sizes, training algorithm, etc., which may all affect the final ML model. Algorithmic transparency relates to the uniqueness of the final solution. Is the result reproducible even if all design choices are reported thoroughly? Oftentimes, there are several local minima where a model can get stuck and the result of two training sessions can end up quite different. If that is the case, then the algorithm is not transparent.

\paragraph{Interpretability}
Interpretability is often what is meant with explaining a model. Interpretability is about describing the properties of a ML model to a person, i.e. “the mapping of an abstract concept (e.g., a predicted class) into a domain that the human can make sense of” \citep{MontavonSM2018}. Methods for interpretability try to determine and show which input data (or part of the input data) that was responsible for the model prediction. Some models are intrinsically interpretable, e.g. linear models, and can be preferred even though they may be less accurate. Methods for interpreting ML models include the SHAP technique, and there are many others.  If the output (decision) from a ML model can be described locally, e.g., by heatmaps, filter responses, local expansions, etc., then one can say that the model is interpretable.

\paragraph{Explainability}
Explainability is a collection of interpretations with further contextual information. It deals with causality, the “what”, “how”, and “why” questions \citep{Miller2019}. It may not be enough to look at a single data point and an interpretation of the resulting prediction to explain a model, e.g., which pixels in an image were important for a prediction. Knowledge creation for scientific purposes requires an understanding of the relationships encoded in the model, using concepts that are understood by the scientific community, and agreeing with (and using) prior domain knowledge. This often means simplifying; can a (large) group of features be collected together into one or a few concepts, can relations be expressed with few and simple operations, and so on?
With this definition of Explainability, it is clear that much work remains to achieve explainable ML models. Essentially all work so far has focused on interpretability.

\paragraph{Domain Knowledge}
The term domain knowledge refers to the background knowledge of the field or environment to which the methods are being applied. Three aspects are involved according to Rueden et al. \citep{RuedenEtAl2021}: type of knowledge, representation and transformation of knowledge and integration of the above into the ML approach. They arrange different types of knowledge along a continuum from sciences, to engineering towards individuals’ intuition. Domain knowledge can be integrated into an ML approach in the training data, in the hypothesis, the training algorithm and the final model. 

\subsection{XAI for images}
Given the success of deep learning models for images analysis there is a large focus on explainable methods for such models. The recent success of deep learning models for diagnostic purposes within the healthcare domain has also put a focus on the requirement of explainability. There exist several review papers on explainable deep learning models in medical image analysis, such as \citep{SinghSL2020} and \citep{GulumTK2021}.

A common approach to explain a deep learning imaging model is to create attribution maps. The attribution values can be interpreted as the contribution or relevance of each input feature for the given task. In case of images where input features are the actual pixel values, attribution maps are often presented as an image of the same size as the input images and provides a direct visual explanation of the method. Many attribution methods can be applied to trained deep learning models, such as the family of convolutional neural networks, without any modifications of the underlying architectures or learning procedures. This makes them ideal for many applications also outside of the medical domain such as satellite imaging.

Attribution maps can be created using two broad approaches, perturbation (e.g. occlusion) based methods or backpropagation (e.g. gradient) based methods. The former approach is modifying the input image and measuring the effect it has on the output of the model. The perturbation can be modifying individual pixel values or be larger by masking out larger patches of the input image (occlusion) \citep{ZeilerEtAl2014}. It can highlight both negative and positive responses. A drawback is the computational requirement of perturbation methods and it may be difficult to determine suitable perturbation schemes.

Backpropagation based methods for attribution maps are model specific and follow various responses both forward and backward through the convolutional layers of the network. Methods that compute gradients fall under this category, such as the GradCam family \citep{SelvarajuEtAl2017}. Here we also find DeepLift \citep{ShrikumarGK2017} and Deep SHAFT \citep{ChenLL2019}, where the latter is using game theory based SHAP values but modified to work for deep learning models. Many backpropagation methods can produce attribution maps for both positive and negative contributions, such as the DeepLIFT method.

\section{Methods}
The available literature in this research domain is presently not very large but growing rapidly. We conducted a review of 30 papers in which machine learning is used on satellite data and images to predict development indicators or poverty. The paper selection process was straightforward and involved inventorying arxiv.org database where a significant share of the papers in this field are published. Additionally, Google Scholar searches, existing reviews and references in references were harvested for relevant papers. 

As far as possible the structure provided by Roscher et al. \citep{RoscherEtAl2020} is used to organize the review. The presented models and analyses are evaluated with respect to the seven aspects, described by Roscher et al. \citep{RoscherEtAl2020}, as listed and described in the tables below: model transparency, design transparency, algorithmic transparency, interpretability, algorithmic explainability, using domain knowledge for feature design, using domain knowledge for generating the hypothesis, using domain knowledge in the training (loss function), and applying domain knowledge to the final model. How much the work in each paper fulfills these aspects is grouped into three levels: well, somewhat, and poorly. In the figures, the three levels are represented with the three colors green, yellow, and red, respectively.

\begin{table}
	\caption{Transparency}
	\centering
	\begin{tabularx}{\textwidth}{XXX}
		\hline
		Model transparency     &   Design transparency   & Algorithmic transparency \\
		\hline
		 The model is mathematically transparent (eg., functions, size, deterministic,…) & Decisions in the design (eg., kernels, layers, units, …) are clearly described and motivated &  The way to find the solution is "unique", the solution can be found again (eg., stopping criteria…) \\
		 \hline
		 The paper should describe the model structure so that the model can be written down mathematically by an expert. If this is the case then it is valued as “mathematically transparent” (green). If the paper mentions the use of a method but nothing on what the final model looks like, then it is valued as “not mathematically transparent” (red). If the paper is somewhere in between those two cases, e.g., that the feature extraction would be hard to reproduce but the remaining model is ok, then it is valued as “somewhat mathematically transparent” (yellow).  
		 &  
		 The paper should describe the model design choices well enough so that an expert can understand and repeat those choices. If this is the case then the model is valued as “design transparent” (green). If no motivation or specification is given on the design choices, then the model is valued as “not design transparent” (red). If the description in the paper is in between those two, some choices are described well, others not so well, then the paper is valued as “somewhat design transparent” (yellow).  
		 & 
		 In ML there is an acceptable level of non-uniqueness in the solution. Parameters may not be identical but the functional result of repeated training can be almost identical. If the algorithm and training method used in the paper are described such that repeated runs should produce a functionally almost identical result, then the paper is labeled “algorithmically transparent” (green). If important issues, like stopping criteria, are not described, then the paper is valued as “not algorithmically transparent” (red). Anything in between is valued “somewhat algorithmically transparent” (yellow). 
		 \\
		\hline
	\end{tabularx}
	\label{tab:table_trans}
\end{table}

\begin{table}
	\caption{Interpretability and Explainability}
	\centering
	\begin{tabularx}{\textwidth}{XX}
		\hline
		Interpretability    & Algorithmic explainability  \\
		\hline
		 The properties of the final model are described in a way understandable to a human.
		 & 
		 What, how and why? 
		 \\
		 \hline
		 The paper should make an effort at describing the properties of the model in an understandable way. In the case of a linear model, this is straightforward. For simpler decision trees, this is also straightforward. For more complicated models, saliency maps or heatmaps can be used to illustrate what the model reacts to. If such a description is in the paper and well explained, then it is valued as an “interpretable” model (green). If there is no attempt at all at describing the properties of the model, then it is valued as “uninterpretable” (red). Cases between these extremes are valued as “partly interpretable” (yellow).
		 &  
		 Does the paper make an attempt at explaining why a certain prediction is made? For example, why are some villages considered poor and others not? What would need to change in the satellite data for a village to move from poor to not so poor, etc,? If the paper makes an attempt at this, or if the answer is obvious from the model structure (eg. a linear model), then it is valued as “explainable” (green). If there is no discussion at all in the paper on this and the model is not straightforward to explain, then it is valued as “not explainable” (red). Cases between these two are valued as “partly explainable” (yellow).
		  \\
		\hline
	\end{tabularx}
	\label{tab:table_inter}
\end{table}

\begin{table}
	\caption{Domain Knowledge}
	\centering
	\begin{tabularx}{\textwidth}{XXXX}
		\hline
		Data (features) & Hypothesis (model structure) & Training (loss function) & Final model (constraints)  \\
		 \hline
		 Does the paper use domain knowledge in the selection of features, or when engineering the features? If this is the case, then it is valued as “domain knowledge in data” (green). If the work relies completely on learning from data without any prior knowledge, then it is valued as “no domain knowledge in data” (red). Cases in between, eg. using both domain knowledge features and pure learning from data, are labeled as “some domain knowledge in data” (yellow). 
		 &  
		 Is there some hypothesis built into the model? For example certain symmetries, or that some features have positive or negative impact on the prediction? If this is the case, then the work is valued as “hypothesis included in model” (green). If there is no such hypothesis, then it is valued as “no hypothesis included in model” (red). In unclear cases, or if there is some weak hypothesis used, then it is valued as “some hypothesis included in model” (yellow).
		 &
		 Does the training procedure utilize domain knowledge? For example, that the loss function should have a certain form specific for this problem. If this is the case, then the work is valued as “domain knowledge in training” (green). If a standard loss function and no domain knowledge is used in the training, then it is valued as “no domain knowledge in training” (red). In unclear cases then it is valued as “some domain knowledge in the training” (yellow).
		 &
		 Is it checked if the final model fulfills known, or expected, relationships or constraints? The simplest example can be that development index values should be positive. Others could be that strong increases in predicted values should be manifested by certain changes in the features. If there is a proper discussion on this in the paper, then it is valued as “domain knowledge checked in final model output” (green). If there is no such discussion in the paper, then it is valued as “no domain knowledge checked in final model output” (red). In cases in between these two, it is valued  as “some domain knowledge checked in final model output” (yellow).
		 \\
		\hline
	\end{tabularx}
	\label{tab:table_domain}
\end{table}

The categorization involves estimating boundary cases and can be illustrated with a few examples. One is the recent study by Lee \& Braithwaite \citep{LeeB2020}. They use two machine learning models in their work: XGBoost and Convolutional Neural Networks. Both result in well-defined deterministic functions. The parameters for training them are provided in the paper and a researcher proficient in these methods should be able to reproduce them. Their work is labelled as “green” for all three parts of transparency, although one could argue that it should perhaps be “yellow” for the third part (Algorithmic transparency). For the input to their models they use well-defined sources, but on the output side (the label side) they both correct survey locations by hand and use a different wealth index than others. Therefore, their data should be made available for the work to be perfectly repeatable, but there are ethical aspects that need to be considered. It is also unclear if their final prediction is a combination of the feature-based and the image-based models or if it is the output of only one of them. However, all their final models (parameters) should be possible to share without ethical issues, and thus made available for others to explore for explainability, so their models are labelled as transparent. Lee \& Braithwaite \citep{LeeB2020} do not discuss how different features affect the output or what needs to be changed in a image for the output to change, so their work is labelled “red” for both interpretability and explainability. However, in their work they include features believed to be important for the prediction so we label it as “green” for domain knowledge features. Domain knowledge is not used in other ways for the model building, e.g. for the cost function or to constrain the output so the rest of the domain knowledge aspects are labelled “red”.

Another example is the study by Yeh et al. \citep{YehEtAl2020a} where deep learning models are used to estimate asset wealth across approximately 20,000 African villages. The methods are well described from a transparency point of view, including details of obtaining the satellite data, good specification of the deep learning models and a good description of the model selection procedure such that it can be reproduced by other researchers. To further strengthen the transparency both data and code are available for download.

For interpretability some studies use models that are inherently interpretable, such as linear models (e.g. Njuguna \& McSharry \citep{NjugunaM2017}) or simpler decision trees (e.g. Watmough et al. \citep{WatmoughEtAl2019}), but for deep learning models this is more challenging. A straightforward approach for models based on convolutional neural networks is however to visualize learned filters as a form of interpretability (e.g. Xie et al. \citep{XieEtAl2016}, Tan et al. \citep{TanEtAl2020}, Pandey et al. \citep{PandeyAK2018}).

Head et al. \citep{HeadEtAl2017} present an interesting study where they model wealth and other indicators (for example child weight and water accessibility) that could be related to wealth. They find that whereas wealth can be well modeled, it is not as straightforward with the other indicators, which is surprising. This represents an exceptional case where domain knowledge is used to check if the models exhibit expected relationships. 

\section{Results from the review}
In our review, “machine learning” includes methods like decision trees, random forest, boosting trees, support vector machines, nearest neighbor clustering, neural networks, deep neural networks, Gaussian processes, and Bayesian models. Three cases with linear models are also included. The reviewed papers are listed in Table \ref{table:summary} with comments on the data and models used.

The work presented in each paper was evaluated with respect to the nine aspects listed above, and there were both clear and borderline cases as illustrated in the examples above. The results should not be interpreted for each paper individually but as a result for the group of papers in this field.

\begin{figure}
	\centering
	\includegraphics[scale=0.75]{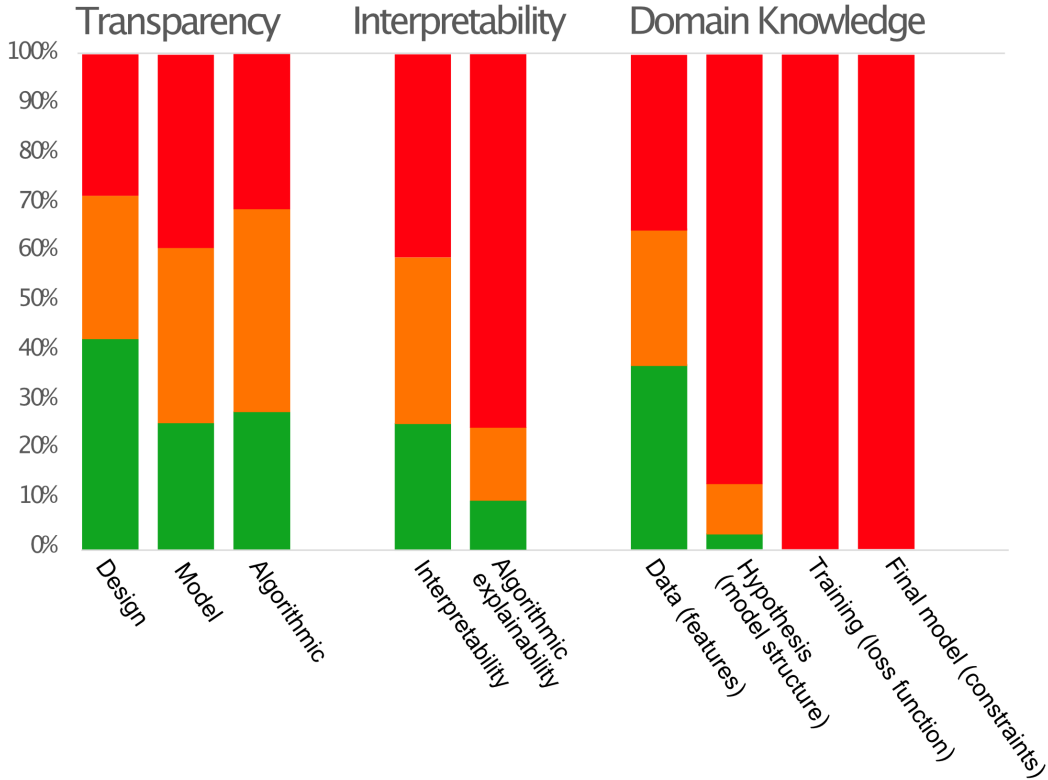}
	\caption{Fractions of the reviewed papers that were classified as transparent (green), not transparent (red), and partly transparent (orange) with respect to the models, the design and the algorithms used.}
	\label{fig:trans}
\end{figure}

The first part of the seven aspects deal with the transparency of the approaches. This relates to how well the work is documented, if the models can be repeated, and if the final models can be written down as functions in mathematical form. As Figure \ref{fig:trans} shows, many papers do this well. However, far from all papers are written such that the work could be reproduced. The second part, interpretability and explainability, is a weak part in this field (and in many other fields too). Figure \ref{fig:trans} shows that few researchers attempt to interpret their models, or even to illustrate what data that leads to certain predictions. The explainability is even weaker; the models that are explainable tend to be simple decision trees or linear models. Very few researchers approach the issue of explaining the model prediction.

The third part of the seven aspects deal with domain knowledge; is domain knowledge used, e.g., to build the models, to select features, or to check if the final models satisfies expected constraints or behaviors. The use of domain knowledge for feature selection is common in the papers dealing with feature-based models. However, domain knowledge is not commonly used in other aspects of the modeling. 

\begin{landscape}
\footnotesize
\begin{center}
\begin{longtable}{|p{1cm}|p{1cm}|p{8cm}|p{11.5cm}|}
  \caption{A list of all reviewed papers including comments on methods and data used.} \label{table:summary}\\
  \hline 
  Ref. & Year & Method & Data \\ 
  \hline
\citep{AngeliniEtAl2019} &
2019 &
Support Vector Machines (SVM) and k-Nearest-Neighbors (kNN).
&
Input: Several computed features from low resolution Sentinel-2 satellite images.
Output: Publically available Afrobarometer survey responses for Kenya, Nigeria, and Tunisia. Infrastructure, living conditions, perception of government corruption.
\\ \hline
\citep{AyushEtAl2020}
&
2020
&
Gradient descent boosting trees, but using deep neural networks (Yolov3) for object detection in the feature generation step.
&
Input: xView (DigitalGlobe) high resolution satellite images. Features are extracted from these using Yolov3.
Target: Living Standards Measurement Study for Uganda. Consumption expenditure.
\\ \hline
\citep{AyushEtAl2021}
&
2021
&
Gradient descent boosting trees. Deep neural networks (Yolov3) for object detection in the feature generation step. Reinforcement learning method for selecting low-resolution or high-resolution images.
&
Input: xView (DigitalGlobe) high resolution satellite images and low resolution Sentinel-2 satellite images.
Target: Living Standards Measurement Study for Uganda. Consumption expenditure.
\\ \hline
\citep{BabenkoEtAl2017}
&
2017
&
GoogleNet Convolutional Neural Network(CNN) model. Trained directly to predict poverty.
&
Input: Digital Globe and Planet satellite imagery (RGB). High res and low res.
Target: Survey data from the 2014 MCS-ENIGH. Income per adult equivalent. Three poverty groups (fraction of households living in poverty). Poverty lines.
\\ \hline
\citep{Chen2016}
&
2016
&
First, ResNet50 Convolutional Neural Network (CNN) model trained on ImageNet.
The second step is done with linear ridge regression.
&
Input: First, 8 bands from year-averaged landsat-7 satellite images. Second, the features from the CNN.
Target: First, night time lights. Second, a normalized wealth score computed from survey data collected by the World Bank in the Living Standard Measurement Study (LSMS).
\\ \hline
\citep{EngstromHN2021}
&
2021
&
Features are identified using a combination of Convolutional Neural Network (CNN) and classification of spectral and textural characteristics. In some cases (roads) is manual identification used. 
Finally a linear LASSO model that uses the features as input.
&
Input: Object and texture features derived from HSRI (High Spatial Resolution Imagery). 
Target: Household estimates of per capita consumption imputed into the 2011 Census of population and housing  (Sri Lanka).
\\ \hline
\citep{EngstromEtAl2019}
&
2019
&
Linear regression (LASSO) for selecting the 15 most important features. Random Forest regression for the poverty index.
&
Input: Quickbird-2 multispectral (Blue, Green, Red, and Near-Infrared) satellite images. Several features are computed from these.
Target: The Ghana Living Standards Survey Round Six (2012-2013). The 2010 Population and Housing Census gathered information from each household on September 26, 2010 (Accra). Neighborhood poverty rates, slum index.
\\ \hline
\citep{HeadEtAl2017}
&
2017
&
VGG16 Convolutional Neural Network (CNN), pre-trained on ImageNet but fine-tuned with night time lights.
&
Input: Satellite images from Google Static Maps. “Low res”. NOAA nighttime light images from the DMSP-OLS website are used for fine-tuning.
Target: Demographic and Health Surveys (DHS) data. Several indices from these.
\\ \hline
\citep{IrvinLR2017}
&
2017
&
Tries both “standard” Convolutional Neural Network (CNN) and a ResNet CNN network. Tries both training the ResNet from scratch and having one pre-trained on ImageNet data but fine-tuned on their task. Tests with averaging outputs from image tiles, or using more advanced recurrent neural networks for combining (LSTM).
&
Input: Satellite images from Google Maps Static API, zoom level 15.
Target: Demographic and Health Surveys (DHS) data. Poverty prediction with wealth index, split into 4 categories. Malnutrition prediction using height and weight for age scores, split into 6 categories.
\\ \hline
\citep{IrvineWM2017}
&
2017
&
A k-Nearest-Neighbor (kNN) model with k=3. Several features computed from the satellite images.
&
Input: Satellite image data from the National Geospatial-Intelligence Agency (NGA). Commercial data and sometimes four spectral bands. Features (color indices, objects, …) computed from this.
Target: Survey responses collected in sub-Saharan Africa by the Afrobarometer network in its Round 4 (2008/2009) surveys.
\\ \hline
\citep{JeanEtAl2016}
&
2016
&
The VGG-F Convolutional Neural Network (CNN) pre-trained on ImageNet data, fine tuned on night time lights. Ridge regression used for the final model from CNN features to poverty/well-being indices.
&
Input: Satellite image data from Google Static Maps API, zoom level 16 (several zoom levels are tried).
Target: Two are used. Consumption expenditure as measured in the World Bank’s Living Standards Measurement Surveys (LSMS). Expenditures are averaged over clusters and the log expenditure is modeled. Household asset score taken from the Demographic and Health Surveys (DHS).
\\ \hline
\citep{KimEtAl2016}
&
2016
&
ResNet-50 Convolutional Neural Network (CNN) pre-trained on ImageNet and then tuned on night time lights.
&
Input: Satellite image data from Google Static Maps API, zoom level 14, 16 and 18 (all three levels are used).
Target: Asset wealth index computed from the Demographic and Health Surveys (DHS).
\\ \hline
\citep{LeeB2020}
&
2020
&
XGBoost (decision tree) for the feature based model. Convolutional Neural Network (CNN) for the satellite image based model.
&
Input: OpenStreetMap data, the Visible Infrared Imaging Radiometer Suite Day/Night Band (VIIRS DNB) nighttime lights dataset, day-time satellite images (Google Static Map, zoom level 16), and the High-Resolution Settlement Layer (HRSL) datasets
Target: The International Wealth Index (IWI) computed from the Demographic and Health Surveys (DHS) data.
\\ \hline
\citep{LiEtAl2019}
&
2019
&
Gaussian process with Radial Basis Function (RBF) kernel, Stochastic gradient boosting, Partial Least Squares (PLS) regression for generalized linear models, Random Forest (RF), Rotation forest, Support Vector Machine (SVM), Neural network with feature extraction
&
Input: Imagery from version 4 of the 2010 night-time lights by the DMSP/OLS satellites. Several features calculated from these images.
Target: High or low poverty county, as defined by the Chinese government.
\\ \hline
\citep{NiEtAl2020}
&
2021
&
Convolutional Neural Networks (CNNs). Four pre-defined architectures: VGG-Net, Inception-Net, ResNet and DenseNet. The two latter were also modified and tested as two new models.
&
Input: Satellite images from Google Maps Static API. Fine tuning with night-time light imagery from DMSP/OLS satellite.
Target: Demographic and Health Survey (DHS) data. Poverty index. Unclear what type.
\\ \hline  
\citep{NjugunaM2017}
&
2017
&
Linear model of the log(MPI).
&
Input: Four features computed by combining three data sources. Mobile phone call detail records (CDRs). Night-time lights from the National Oceanic and Atmospheric Administration (NOAA). Population density estimates from the Landscan population database produced by US Department of Energy, Oak Ridge National Library.
Target: The multi-dimensional poverty index (MPI) calculated by the National Institute of Statistics in Rwanda (NISR) using data from the 2012 Rwanda Population and Housing Census (RPHC4).
\\ \hline
\citep{PandeyAK2018}
&
2018
&
Convolutional neural network (CNN) architecture for the first task. Trained from scratch. For the second task a multilayer perceptron.
&
Input: First, satellite images from Google Static Maps API, zoom level 16. Images only of villages. Second,  predicted roof material, source of lighting, and source of drinking water.
Target: First, the roof material, source of lighting, and source of drinking water in a region. In a second model, the household income level in a region. The information comes from the 2011 Census of India and Socio-Economic Caste Census of 2011.
\\ \hline
\citep{PerezEtAl2017}
&
2017
&
Convolutional Neural Network (CNN) of type ResNet and VGG-Net. Pre-trained on ImageNet data for RBG-bands, otherwise trained from scratch.
Try both gradient boosted trees and linear ridge regression for predicting AWI from the CNN features, and also (as comparison) directly from night-time light values.
&
Input: First, multispectral satellite imagery from Landsat 7 (several spectral bands tried). Second, CNN features.
Target: First, predicting class of night-time lights (3 classes). Second, Asset Wealth Index (AWI) from Demographic and Health Surveys (DHS) data.
\\ \hline
\citep{PerezEtAl2019}
&
2019
&
Convolutional Neural Network (CNN) with weighted Generative Adversarial Network (WGAN), for the first step. The second step is done with ridge regression.
&
Input: First, multispectral (9 bands) imagery from Landsat 7 (Enhanced Thematic Mapper Plus (ETM+)). RBG images are pan-sharpened (to get 15 m resolution). Second, the features from the CNN.
Target: It is not described fully what they train the CNN to predict, but it includes night-time lights. For a second, linear model, they use Asset Wealth Index (AWI) from Demographic and Health Surveys (DHS) data.
\\ \hline
\citep{PokhriyalJ2017}
&
2017
&
A model with one linear term and one nonlinear Gaussian Process term.
&
Input: Several features extracted from many different data sources; e.g., Open Street Map,  Mobile phone call records (CDRs), World Climate Database, the Digital Elevation Database, Night Time Lights, and the Soil and Terrain Database for Senegal.
Target: The Global Multidimensional Poverty Index (MPI), computed from Demographic and Health Surveys and Multiple Indicator Cluster Survey (DHS-MICS) data.
\\ \hline
\citep{SmithW2018}
&
2018
&
Linear model. The “unlit rural percentage” is used as an indicator for poor rural regions (100 square km regions).
&
Input: Night time lights from Defence Meteorological Satellite Program’s Operational Linescan System (DMSP-OLS). Population estimations from the Oak Ridge National Laboratory LandScan. By combining these is the “unlit rural percentage” (URP) computed.
Target: Comparative Wealth Index (CWI) and Unmet Basic Needs (UBN) from the Demographic and Health Surveys (DHS) data.
\\ \hline
\citep{SteeleEtAl2016}
&
2021
&
Hierarchical Bayesian Geostatistical Models (BGMs). Inputs are features computed from several data sources.
&
Input: Mobile phone call detail records (CDRs) from Telenor, partly for the same households as the market surveys in the input. VIIRS night-time lights. Population counts from CIESIN - Global Rural Urban Mapping Project. Climate indicators from CGIAR-CSI.Vegetation indices from MODIS MOD13A1. Features from Open Street Map. Land cover features from ESA GlobCover Project, IGBP MODIS MCD12Q1, and ONRL DAAC Synergetic Land Cover Product (SYNMAP).
Target: Three development indicators. The Wealth Index from Demographic and Health Surveys (DHS). The Progress out of Poverty Index based on a survey of adults in Bangladesh by InterMedia Financial Inclusion Insight Project in 2014. Two sequential large-scale market research household surveys run by Telenor.
\\ \hline
\citep{SubashKA2018}
&
2018
&
Artificial Neural Network (unspecified architecture). 
&
Input: Night time lights from India Lights API, and GDP values from India statistics.
Target: Poverty estimates from India survey.
\\ \hline
\citep{TanEtAl2020}
&
2020
&
ResNet50 (CNN) for the first task. Ridge regression for the second task.
&
Input: First, Landsat 8 images. Second, the features from the ResNet50 network. 
Target: First, spectral index data (NDVI, MNDWI, and NDBI), and night-time light data (Suomi NPP). Second, development indicators from the Chinese statistical yearbook data.
\\ \hline
\citep{TingzonEtAl2020}
&
2020
&
Pixel-wise modeling. Three models tried: logistic regression, linear support vector machines (SVM), and random forest (RF).
&
Input: Sentinel 2 satellite data, from these were 66 features computed per pixel.
Target: Field data of informal migrant settlements collected by the humanitarian organization iMMAP in 2019.
\\ \hline
\citep{WatmoughEtAl2019}
&
2019
&
Classification decision tree. Image features are computed from the satellite images using a combination of object-based image classification, fuzzy classification, and Random Forest (RF).
&
Input: Features computed from two satellite sources: land use/land cover (LULC) map derived from a QuickBird image, and Moderate Resolution Imaging
Spectro-radiometer (MODIS) data.
Target: A wealth index designed by the authors based on household survey data collected by the authors.
\\ \hline
\citep{WatmoughEtAl2016}
&
2016
&
Random Forest (RF).
&
Input: Satellite data from Landsat Enhanced Thematic Mapper Plus (ETM+) and the MODIS Normalized Difference Vegetation Index (NDVI). Features were computed from these.
Target: A welfare index computed from socioeconomic data at the community level from the 2001 Indian National Population and Household Census.
\\ \hline
\citep{WuT2019}
&
2019
&
First, ResNet50 model pre-trained on ImageNet data, fine tuned on night-time light data. Second, linear ridge regression.
&
Input: First, LANDSAT 8 images. Second, the features from the ResNet50 model.
Target: First, Nighttime light data (Suomi NPP satellite). Second, Gross domestic product (GDP) and total retail sales of consumer goods (TRSCG).
\\ \hline
\citep{XieEtAl2016}
&
2016
&
First, the VGG-F Convolutional Neural Network (CNN) model, pre-trained on ImageNet. Second, logistic regression.
&
Input: First, satellite images from the Google Static Maps API, at zoom level 16. Second, the features from the CNN.
Target: First, night time light intensities from The National Oceanic and Atmospheric Administration (NOAA). Second, data with binary poverty labels from Living Standards Measurement Study (LSMS) survey conducted in Uganda (Uganda Bureau of Statistics 2012).
\\ \hline
\citep{YehEtAl2020a}
&
2020
&
The ResNet-18 Convolutional Neural Network (CNN) model (v2, with preactivation). Pre-trained on ImageNet data. One network for daytime satellite images, one for night-time.
&
Input: Multispectral images from Landsat archives available on Google Earth Engine, and night-time lights images (VIIRS and DMSP).
Target: Wealth index from Demographic and Health Surveys (DHS) data.
\\ \hline
\citep{ZhaoEtAl2019}
&
2019
&
Random Forest (RF). A Convolutional Neural Network (CNN) VGG-F, trained on ImageNet data, was fine-tuned to predict night time light classes (and thus learn features).
&
Input: Night-time lights from the VIIRS Cloud Mask–Outlier Removed (vcm–orm) annual composite NPP-VIIRS DNB data (NOAA/NCEI). Google Static Maps satellite images, zoom level 16 (high resolution). Open Street Map. Land cover maps from the European Space Agency (ESA) Climate Change Initiative. Different features were computed from these.
Target: The Wealth Index (WI) from the Demographic and Health Surveys (DHS).
\\ \hline
\end{longtable}
\end{center}
\end{landscape}

\section{Discussion}
The dramatic success in machine learning over the last decades has led to re-vitalization and progress in unexpected domains. Starting with the seminal papers of Jean and Xie \citep{XieEtAl2016} \citep{JeanEtAl2016} it is now evident that poverty can be accurately estimated from combinations of satellite imagery and machine learning, adding one important step towards the fulfillment of SDGs, particularly those related to poverty reduction.  Specifically, the fundamental questions of “where are the poor and how poor are they” could potentially be answered without launching a new wave of surveys. Recent research has accounted for some of the initial limitations that has been pointed out by Head et al. \citep{HeadEtAl2017} and others, with generalizability being the most important. Lee and Braithwaite \citep{LeeB2020} have shown that it is possible to create a methodology that is generalizable to several countries. While great progress has been achieved in a short time there are several areas that need attention, not least considering reported lack of downstream applications of the methodology. We argue that explainability is essential to support such downstream applications, and explainability means more than just interpretability.
 
In this paper we have discussed how this methodology can be used for scientific discovery, which is asking for a lot, keeping in mind the recency of the research field. As a road map we have used the requirements suggested by Roscher et al. \citep{RoscherEtAl2020}. Our review of the field shows that the status of the three core elements of explainable machine learning (transparency, interpretability and domain knowledge) is varied and not completely fulfilling the requirements set up for scientific insights and discoveries. Transparency matters are often well covered meaning that most of the work is replicable and mathematically sound. Interpretability is not very well covered, but at the same time interpretability is according to the literature where recent efforts are directed and where we found state-of-the art research \citep{AyushEtAl2020} \citep{AyushEtAl2021}. Furthermore, the use of domain knowledge which is important to achieve scientific consistency, is not very well covered in the papers we have assessed. 

The overarching goal is to use satellite data with its unprecedented spatial and temporal coverage to measure various aspects of human welfare. The dominating approach where outputs are benchmarked against survey data has produced some remarkable results but largely overlooks epistemological questions. Information evaluation, specifically what imagery is capable of contributing to this specific domain is not well understood. Head et al. \citep{HeadEtAl2017} concludes that if there is insufficient signal in the image, “no matter how sophisticated our computational model, the model is destined to fail”. Therefore, there are some results that are intuitively inconsistent and difficult to explain. One such example observed by Head et al. \citep{HeadEtAl2017} was the relative under-performance of models designed to predict access to drinking water. They expected the satellite-based features to capture proximity to bodies of water, which in turn might affect access to drinking water. However, an explainable AI approach here could perhaps shed light on this surprising finding.
 
Understanding the interplay of different sensor characteristics (spatial- and spectral resolutions), interactions with different physical environments and the nature of ground truth is crucial. Starting with the latter, the construct of the DHS WI was designed so the survey officer could collect information about assets that were easy to visually observe and register without asking household members. The features in such an asset-based model all come from well-accepted theory suggesting that they represent long-term welfare status and being poor or rich is directly related to how many and what type of assets each household and village possesses. The WI is essentially a look-up-table for “object recognition” but with several objects stored inside houses and therefore inaccessible for remote sensing. Some objects are accessible from above but could still remain undetected for the sensor at hand. With that in mind, it is remarkable that the methodology performs so well, suggesting  unexplained processes may be at play and warrants more research.
 
Satellite imagery is, in many aspects, about what we see is what we get. A rule-of-thumb is that for an object to be observable it should be covered with a minimum of four pixels. Image enhancement techniques and combining spectral information in creative ways can bring out extra detail and high-contrast objects are also more likely to be observable. If we consider the commonly used 2.5 m satellite image (scale level 16 in Google API) imagining it centered over one of the DHS rural villages, what could we expect to observe? The dominating objects would be buildings, roads, agricultural fields, forest patches etc. but also the overall spatial organization of the village. In other words, many features that are not accounted for in WI design. It would be difficult to identify cars (but maybe possible), bicycles, electricity poles, cell phones, television sets, farm animals, etc. although they are all features in the WI. 
 
There are grounds to suspect that the satellite image adds poverty related information beyond what is accounted for in the WI but still correlated to overall poverty. It is plausible that this information is related to the environment and spatial organization of villages which is known to be powerful determinants of poverty. Knowing what features, or combinations of features in the satellite image (at different resolutions) correlates with wealth is crucial but is also a potential for scientific discovery. Is it consistent with domain knowledge? Having this information would shed light on the important relation between geographically determined poverty and other forms of poverty. The recent studies by Ayush et al. \citep{AyushEtAl2020} \citep{AyushEtAl2021} represent interesting starting points. Here, as a first step, objects were detected in the images using object detection methods. A feature vector was then constructed using the counts of different objects, augmented with confidence and size estimates. These feature vectors were then used to model a poverty index, using a tree based method. This approach provides an increased level of explainability due to features that are directlty interpretable and models that inherently provide some explanations of how features affect the outcome. There is still room for improvement in terms of expanding the set of objects to detect or adding more abstract features, such as landscape characteristics, in addition to further increased model explainability. The advantage of a traditional equation, showing both which terms there are in the equation and what factors there are in front of each term, is obvious; “What does positive change in poor societies look like and how is it achieved?” \citep{Ostberg2018}. We should aim for this simplicity in interpretation, while being generated as part of the machine learning process. 



\bibliographystyle{hplain}
\bibliography{xai}

\end{document}